\documentclass[fleqn,usenatbib]{mnras}

\usepackage{newtxtext,newtxmath}
\usepackage[normalem]{ulem}
\usepackage[T1]{fontenc}
\DeclareRobustCommand{\VAN}[3]{#2}
\let\VANthebibliography\thebibliography
\def\thebibliography{\DeclareRobustCommand{\VAN}[3]{##3}\VANthebibliography}

\usepackage{xcolor}

\usepackage{graphicx}	
\usepackage{amsmath}	
\usepackage {threeparttable}

\title[Aftershocks of SGR 1935+2154 radio pulses]{Similarity to earthquakes again: 
periodic radio pulses of the magnetar SGR 1935+2154
are accompanied by aftershocks like fast radio bursts}

\author[Y. Tsuzuki et al.]{
Yuya Tsuzuki,$^{1}$
Tomonori Totani$^{1,2}$\thanks{E-mail: totani@astron.s.u-tokyo.ac.jp}, 
Chin-Ping Hu$^3$, and Teruaki Enoto$^{4,5}$
\\
$^{1}$Department of Astronomy, the University of Tokyo, 7-3-1 Hongo, Tokyo 113-0033, Japan\\
$^{2}$Research Center for the Early Universe, the University of Tokyo, 7-3-1 Hongo, Tokyo 113-0033, Japan \\
$^3$Department of Physics, National Changhua University of Education, Changhua 50007, Taiwan \\
$^4$Department of Physics, Kyoto University, Kitashirakawa-Oiwake-cho, Sakyo-ku, Kyoto, 
606-8502, Japan \\
$^5$RIKEN Cluster for Pioneering Research, 2-1 Hirosawa, Wako, Saitama, 351-0198, Japan
}

\date{Accepted XXX. Received YYY; in original form ZZZ}

\pubyear{2023}

\begin{document}
\label{firstpage}
\pagerange{\pageref{firstpage}--\pageref{lastpage}}
\maketitle

\begin{abstract}
It was recently discovered that the time correlations of repeating fast radio bursts (FRBs) 
are similar to earthquake aftershocks. 
Motivated by the association between FRBs and magnetars,
here we report correlation function analyses in the time-energy space
for the 563 periodic radio pulses and the 579 X-ray short bursts 
from the magnetar SGR 1935+2154, which is known to have generated FRBs.
Although radio pulses are concentrated near the fixed phase of the rotational cycle, 
we find that when multiple pulses occur within a single cycle, their correlation properties 
(aftershock production probability, aftershock rate decaying in power of time, and more)
are similar to those of extragalactic FRBs and earthquakes.
A possible interpretation is that  
the radio pulses are produced by rupture of the neutron star crust, 
and the first pulse within one cycle is triggered by external force periodically 
exerted on the crust. 
The periodic external force may be from the interaction of the magnetosphere with 
material ejected in an outburst. For X-ray bursts, we found no significant correlation signal,
though correlation on the same time scale as radio pulses may be hidden due to 
the long event duration.
The aftershock similarity between the periodic radio pulsation and FRBs is 
surprising, given that the two are energetically very different, and therefore the energy sources 
would be different.
This suggests that the essence of FRB-like phenomena is starquakes, regardless of the energy source, and it is important to search for FRB-like bursts
from neutron stars with various properties or environments.
\end{abstract}




\section{Introduction}
\label{section:intro}

Fast radio bursts (FRBs) are extragalactic transient objects detected in radio waves 
with millisecond durations, and their source objects and emission mechanisms are 
largely still a mystery, though many theoretical models have been proposed (see
\citealt{Cordes_2019,platts2019living,zhang2020physical,petroff2022fast} for reviews). 
Some FRBs are known to produce recurring bursts, and these are likely to 
originate in neutron stars. In particular, magnetars (highly magnetized neutron stars,
see \citealt{kaspi2017magnetars,esposito2018magnetars} for reviews) have been
considered a promising source of FRBs because of their abundant magnetic energy and 
the bursts of X-rays and gamma-rays that they occasionally induce.
In fact, on April 28, 2020, two extremely bright radio bursts 
(FRB 20200428) similar to 
extragalactic FRBs were detected from the Galactic magnetar SGR 1935+2154, establishing 
that at least some FRBs are generated by magnetars
\citep{andersen2020bright,bochenek2020fast}. 

More than several thousand FRB events have already been detected from 
several extragalactic FRB repeaters, and detailed statistical studies are possible.
An interesting fact already established is that the burst wait-time distribution is bimodal.
Although the long-side peak of the bimodal distribution can be
explained by events occurring randomly by a Poisson process, the origin of the shorter peak has 
not been established \citep{wang2017sgr,oppermann2018non,wang2018frb,Zhang_2018,Zhang_2021,zhang2022fast,zhang2023frbs,li2019statistical,li2021bimodal,gourdji2019sample,wadiasingh2019repeating,oostrum2020repeating,tabor2020frb,aggarwal2021comprehensive,cruces2021repeating,hewitt2022arecibo,xu2022fast,du2023scaling,jahns2023frb,10.1093/mnras/stad1739,Wang_2023}.
In the previous study, we analyzed the two-point correlation function in the two-dimensional space 
of occurrence time and energy of repeating FRBs, and showed that the statistical 
characteristics of FRBs are remarkably similar to those of earthquakes but different from 
solar flares \citep[][ hereafter TT23]{Totani+23}.  
The similarities between FRBs and earthquakes can be listed as follows: (1) the probability 
of a single event followed by correlated aftershocks is about 10-50\%, (2) the
aftershock rate follows the Omori-Utsu law \citep{omori1895after,Utsu1957,Utsu1961}, 
which decays as a power of event
time interval $\Delta t$, (3) the power law extends toward shorter $\Delta t$
to the typical durations of FRBs and earthquakes, (4) even if the average event 
rate changes significantly due to fluctuations in activity,
the aftershock rate (or probability)
is universal and stable among different FRB sources, 
and (5) there is little correlation between the aftershock energies.
These results suggest that FRBs are caused by ruptures of solid crusts 
at neutron star surfaces.

Then it is an interesting question whether a similar time correlation can be found in 
magnetar bursting phenomena. Thousands of
short bursts from magnetars have been recorded
in X-rays with a duration of about 0.01--10 seconds, which are one of the prominent
features of magnetars \citep{kaspi2017magnetars,esposito2018magnetars}. 
SGR 1935+2154 has been one of the most active magnetars this decade, exhibiting 
several X-ray outbursts. 
These outbursts are characterized by a significant increase in persistent X-ray flux, 
typically by 
a factor of 10--1000, and can last from months to years. 
During its April 2020 outburst, NICER recorded hundreds of short X-ray bursts just half 
a day before FRB 20200428 \citep{younes2020nicer}. 
In October 2022, a similar pattern of bursting activity was observed, 
followed a few hours later by the detection of a bright radio burst 
\citep{Giri_2023, Maan_2022, Hu_2024}.

Radio pulsations are also detected from several magnetars. 
These radio pulsations are a transient emission and appear in association with an 
X-ray outburst.
The recently reported radio pulsation detected from SGR 1935+2154, which is the sixth member of
the radio magnetar population, about half a year 
after FRB 20200428 is particularly interesting because it exhibits features 
(narrow-band emissions and frequency drifts)
similar to FRBs, though much fainter \citep{zhu+23,wang2023atypical}.

Therefore, here we perform the same two-point correlation function analysis as 
TT23 on the periodic radio pulses \citep{zhu+23,wang2023atypical}
and X-ray short bursts of SGR 1935+2154 in 2020 and 2022 \citep{younes2020nicer, Hu_2024}
to investigate the nature of time-energy correlation about bursting phenomena in magnetars. 
The magnetar SGR 1935+2154 rotates with a period of $P = 3.24$ s and has a spin-down rate of 
$\dot{P} = 1.43 \times 10^{-11}\mathrm{ss}^{-1}$.
The important physical quantities derived from these are 
the surface dipole magnetic field strength of ~$2.2\times 10^{14} \ \mathrm{G}$,
the characteristic age of about $3.6 \ \mathrm{kyr}$, and the
spin-down luminosity of $1.7\times 10^{34} \ \mathrm{erg} \,\mathrm{s}^{-1}$ \citep{israel2016discovery}. Although the distance to SGR 1935+2154 
is highly uncertain \citep{israel2016discovery,Surnis_2016,zhou2020revisiting}, here we adopt 6.6 kpc following \citet{zhou2020revisiting,zhu+23}.

\section{The Data}

\subsection{Radio pulses}

The radio pulses from SGR 1935+2154 have a short pulse width of about 1 ms and 
are concentrated at the specific phase of the rotation cycle \citep{zhu+23}.
We use for our analysis the 563 pulses detected by
the Five-hundred-meter Aperture Radio Telescope (FAST) during the total observation time of 
17 hours from 9 October to 7 November 2020, as reported in \citet[][hereafter
referred to as the R20 data set, see also Table \ref{table:radio and X-ray}]{wang2023atypical}. 
These pulses were detected at 464 rotational cycles among the $1.9 \times 10^4$
cycles during the total observation time, and hence the probability of 
detecting a radio pulse per cycle is about 2.5\%.
The radio pulses are narrow-band and showing frequency-drifts, which are rarely observed
in normal pulsars, but rather commonly seen in extragalactic FRBs, although the 
average fluence of SGR 1935+2154
radio pulsations are 7--8 orders of magnitude lower than that of FRB 20200428. 
We use the solar system barycentric time $t$ of the pulse peak arrival
and emitted energy $E$ of pulses
from the table given in the original paper \citep{wang2023atypical}. 
The distance to SGR 1935+2154 assumed in \citet{wang2023atypical} is the same
as that we adopt in this work ($6.6 \ \mathrm{kpc}$),
but it should be noted that our correlation function analysis is not affected by 
the distance uncertainty because we look only the logarithmic energy difference
between a pair of two events. 
Observations were made over 14 runs on 13 days, and the start and end times for each run 
are taken from the table in \citet{wang2023atypical}. 

To examine the relationship between activity levels and the nature of the correlations, we divided this sample into three sub-samples. The pulse detection rate was particularly high
during the first three days (MJD 59131.4--59134.4)
of the entire observation period by 
\citet[][see their Fig. 4]{wang2023atypical}, and we call this the high-rate sample. 
During the following 17 days, activity was low and the pulse detection rate was less 
than 1/10 of the high-rate sample. We call this the low-rate sample.
Activity increased again during the last two days, with a pulse detection rate 
about half that of the high-rate sample. We call this the mid-rate sample. 
These sub-samples are also summarized in Table \ref{table:radio and X-ray}.

\subsection{X-ray bursts}

We analyze two samples for NICER X-ray bursts (Table \ref{table:radio and X-ray}). 
The first sample (the X20 sample
hereafter), reported by 
\citet{younes2020nicer}, is bursts detected on April 28, 2020 
(14 hrs before FRB 20200428).
This period is the most intense burst active episode observed to date for SGR 1935+2154.
We use 217 bursts detected in the first good time interval (GTI) 
whose duration is 1120 seconds for correlation analysis, 
omitting 6 bursts detected in later GTIs because of the small number. 
Energy flux $F$ is not reported for 12 bursts in the 217 bursts of the first GTI,
and hence these are dropped from the final sample. 
The total energy of a burst is estimated as $E=4\pi D^2 \, F \, T_{90}$
where $D = 6.6$ kpc is the distance and 
$T_{90}$ is the burst duration estimated as the time interval during
which 5--95\% of the burst fluence is accumulated. 

The second data set is 378 short X-ray bursts detected by NICER from October 12, 
2022 to October 18, 2022 \citep[][the X22 sample hereafter]{Hu_2024}. 
First, we identified burst candidates by the Bayesian-block analysis \citep{Scargle_2013}.
Subsequently, we calculate the chance probability that the total counts in each burst 
represent a Poisson random fluctuation, considering the count rate of approximately 3 
seconds before and after the block.
Bursts with a probability lower than $3\times 10^{-7}$, corresponding to a detection 
significance of 5$\sigma$, are classified as confirmed bursts. 
The total observation time during this period is 8.25 hr, and the mean event rate
is about 10 times lower than the X20 sample. The total number of photons in $T_{90}$ of a burst
given in this data set is used as the proxy of the total energy in the
correlation function analysis. As explained later in Section \ref{section:methods}, 
the entire observation period is divided into sub-periods (each sub-period within
one orbit of NICER, which is different from
GTIs), and the number of pairs for 
correlation function calculation is counted only within each sub-period.
Thus, if there is only one event in a sub-period, it is excluded from the correlation 
function analysis. Since 15 events are excluded for this reason, 
the final number of events in the X22 data set is 374.

\begin{table*}
	\caption{The pulse/burst samples of SGR 1935+2154 used in this study.}
    \begin{center}
	\label{table:radio and X-ray}
    \hspace{-1.3cm}
	\begin{tabular}{lccccccccc} 
		\hline
		Sample name& Telescope & Period (MJD)&Observation time (hours)&Sub-periods&Events&Mean event rate$^a$
  ($\rm{hours}^{-1}$)\\
        \hline
        \hline
        R20 & FAST & 59131.38-59152.49 & 17 & 13 & 563 & 70.54\\ 
        R20 (high rate) & & & 3 & 3 & 353 & 124.76\\
        R20 (mid rate) & & & 2 & 2 & 99 & 52.72\\
        R20 (low rate) & & & 12 & 8 & 111 & 9.59\\
		X20 & NICER & 58967.03-58967.04 & 0.31 & 1 & 205 & 658.99\\  
        X22 & NICER & 59864.73-59869.53 & 8.70 & 26 & 374 & 50.96\\ \hline
	\end{tabular}         
    \end{center}
    \flushleft{
    {\footnotesize 
    $^a$The mean event rates averaged over all sub-periods, with a weight 
    by the number of events ($N_{\rm ev}$).} }
\end{table*}

\section{Correlation function analysis}

\subsection{Methods of correlation function calculations}
\label{section:methods}
The method for calculating the two-point correlation function is mostly the same as in TT23, and here we describe briefly focusing on the different points
from the previous study. We compute the two-point correlation function $\xi$ in the 
two-dimensional space of $\Delta t$ and $\Delta \lg E$, where
$\Delta t \equiv t_2 - t_1$ ($t_2>t_1$) is the time difference between a pair
of two events (at time $t_1$ and $t_2$ with energies $E_1$ and $E_2$)
and  $\Delta \lg E \equiv \mathrm{lg}\;E_2-\mathrm{lg}\;E_1$ 
is the logarithmic energy difference ($\mathrm{lg} \equiv \mathrm{log}_{10}$). 
The function $\xi (\Delta t, \Delta \lg E)$ is defined as the excess of the number density of pairs
compared with the uncorrelated case, and hence 
the number of pairs ($dN_p$) in a bin at ($\Delta t$, $\Delta \lg E$) can be written
as
\begin{equation}
    dN_p = (1+\xi)\,\overline{n}_p\, d(\Delta t)\, d(\Delta \mathrm{lg}\,E) \ ,
\end{equation}
where $\overline{n}_p$ is the expected pair number density in the uncorrelated case.

To estimate $\overline{n}_p$
we need to generate random and uncorrelated bursts by the Monte-Carlo method. 
Random data were generated by a Poisson process assuming a constant event rate and energy distribution, and the energy distribution was constructed empirically from the data sets,
as in TT23.
To reduce statistical error, the random data was generated with the sample size
100 times larger than that of the real data. 
We use the Landy-Szalay (LS) estimator \citep{Landy+1993},
\begin{equation}
    \xi(\Delta t, \Delta \mathrm{lg}\,E) = \frac{DD\,-\,2DR\,+\,RR}{RR} \ , 
\end{equation}
for the time correlation function $\xi(\Delta t)$, because this estimator is known
to have a small variance from the right value. Here, 
$DD$ and $RR$ are the number of pairs in the real and random samples 
(normalized by their different sample sizes), respectively,
in a given bin of the $\Delta t$-$\Delta \mathrm{lg}\,E$ space, and 
$DR$ is the number of cross-pairs between the two samples. 
However, for the two-dimensional correlation function, 
the natural estimator $DD/RR - 1$ 
was used, because in bins with small $DD$, the LS estimator sometimes 
produces unphysical results of negative $1 + \xi$.

Poisson statistics about the number of pairs
can most simply estimate the error in the correlation function, 
but it is known to underestimate if $\xi$ is not zero.
We also tried the jackknife method in our previous work, but jackknife errors were not 
much different from the Poisson errors, and the jackknife errors suffer from a large
uncertainty because of the small sample sizes. Therefore, in this study, we will  
use the Poisson error and not consider the correlation of errors between different bins.
This treatment is sufficient for this study, which does not require rigorous estimation of 
parameter errors.
Poisson statistical errors for a small number of pairs (close to 1) 
were calculated according to \citet{Gehrels1986}.

Radio observations from the ground can be continuous for only a few hours at most during a day.
Following TT23, the entire observation period is divided into sub-periods, with one 
sub-period consisting of data from the same observation date.
We consider only pairs within each sub-period, and hence 
sub-periods in which only one event occurs are not used in our analysis 
because of no available event pairs. For random data generation, the event rate and energy distribution 
were held constant within each sub-period. 
If there is a gap in the middle of a day's observations, the correlation function was 
calculated by not generating random data events during the gap, using the
information on the start and end times of each observing run.
In the case of NICER X-ray data, a continuous observation is at most for 2000 s
since the orbital period of the International Space Station is 90 minutes. 
Therefore, a sub-period is defined as a period of data taken within one orbit. 
There are many short interruptions of observations between GTIs
within a single sub-period, and
these were taken into account in the analysis by referring to the GTI
information of NICER and by not generating random data during interruptions.

\subsubsection{A special treatment for radio pulsation data}

\begin{figure*}
	\includegraphics[width=160mm]{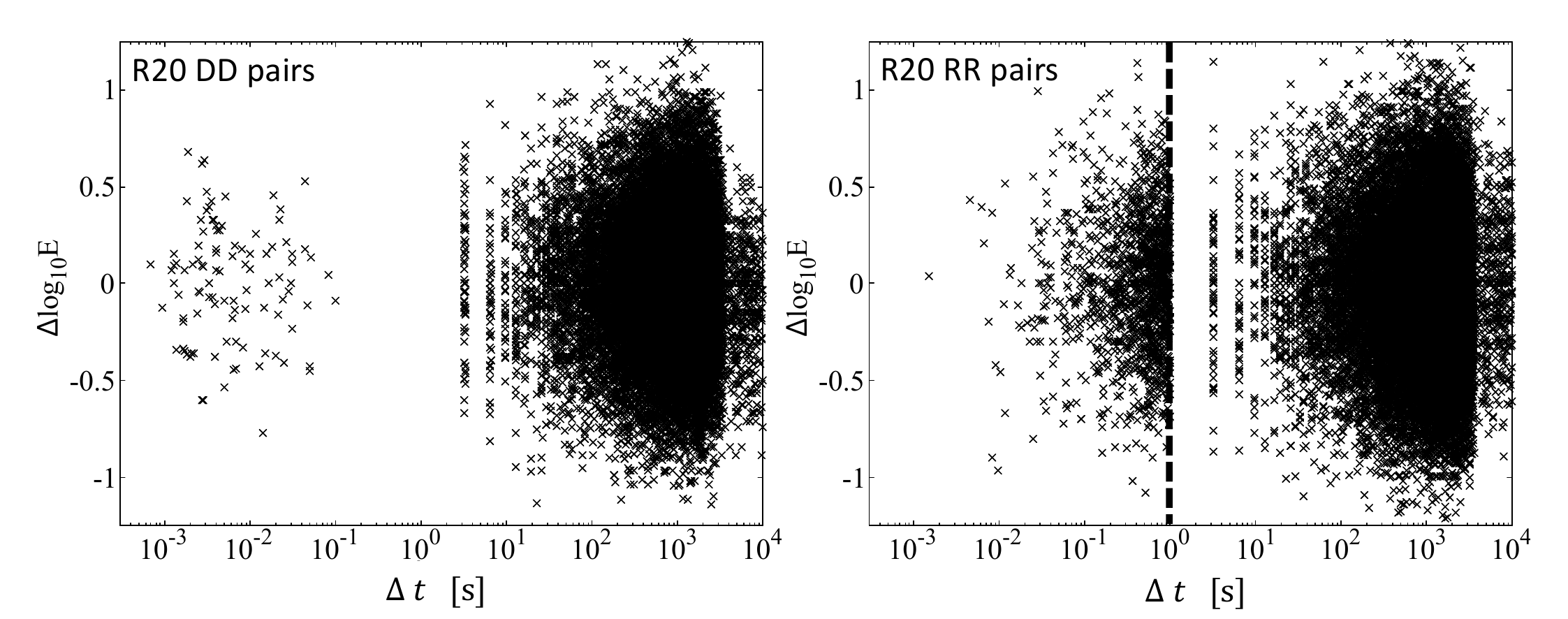}
    \caption{Distribution of DD (left) and RR (right) pairs of the SGR 1935+2154 radio pulses 
    (the R20 data set). Since the method of generating RR pairs is different for 
    $\Delta t > 1$ s and $<$ 1 s, they are shown separately.
    To make the distribution easier to see, we generated the random data 
    with $N_{\rm ran}$ = 10 times more than the actual data in the region of
    $\Delta t < 1$ s, while for $\Delta t > 1$ s we set $N_{\rm ran} = 1$. }
    \label{fig:DDandRR}
\end{figure*}

As reported in \citet{zhu+23}, radio pulses are emitted only at a specific phase within the 3.24 s period.
Therefore, as shown in Fig.\ref{fig:DDandRR}, $\Delta t$ of the 
DD pairs are distributed only at integer multiples of the period $P$ 
in the region of $\Delta t > 1$ s, while pairs of $\Delta t < 0.1$ s appear
because occasionally multiple pulses are emitted at intervals of 0.1 s or less 
within a single cycle. Therefore the random samples were generated separately
for the two regimes, as follows (see Fig.\ref{fig:DDandRR} right panel). 
In $\Delta t > 1$ s, random events were assumed to occur 
randomly on grids separated by the period $P$, allowing only one event on a grid. 
Then, in this region, the correlation 
analysis would examine the presence or absence of correlation compared to the hypothesis 
that the pulses occur randomly with a certain probability in a rotation cycle.
On the other hand, for $\Delta t < 1$ s, the random events were assumed to occur randomly 
in a continuous and uniform time distribution, not on grids, at the observed event
rate averaged over a sub-period. 
This allows us to verify whether correlated aftershocks of $\Delta t < 1$ s
occur following a given radio pulse, as found in TT23 for extragalactic FRBs.

\subsection{Results}

\subsubsection{Radio pulses}

\begin{figure}
	\includegraphics[width=\columnwidth]{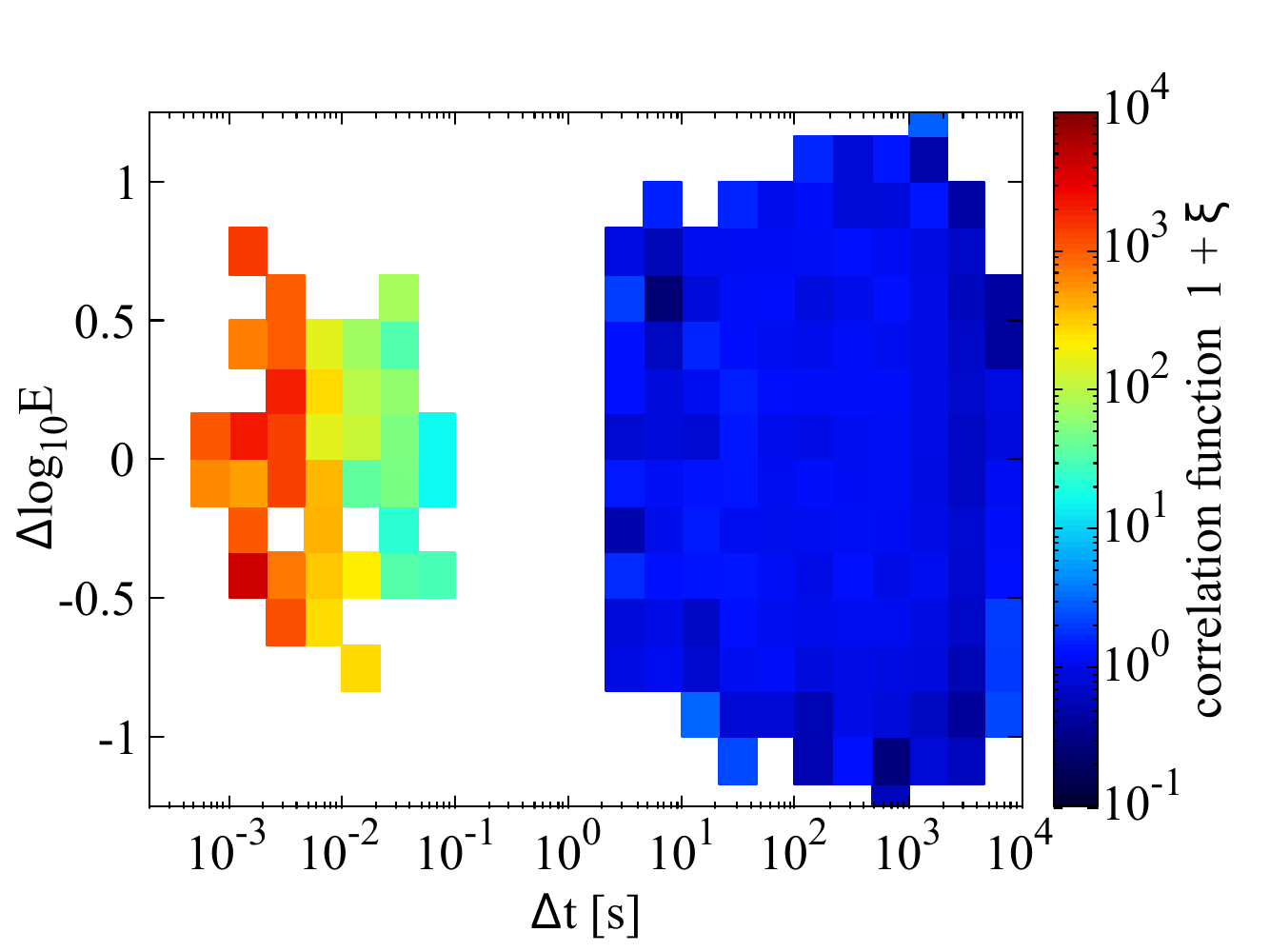}
    \caption{The two-dimensional correlation function $\xi$ in the $\Delta t$-$\Delta \lg E$
    space of the radio pulses of SGR 1935+2154 (the R20 sample). Bins with zero DD pairs 
    are not colored.}
    \label{fig:2Dxi}
\end{figure}

\begin{figure}
	\includegraphics[width=\columnwidth]{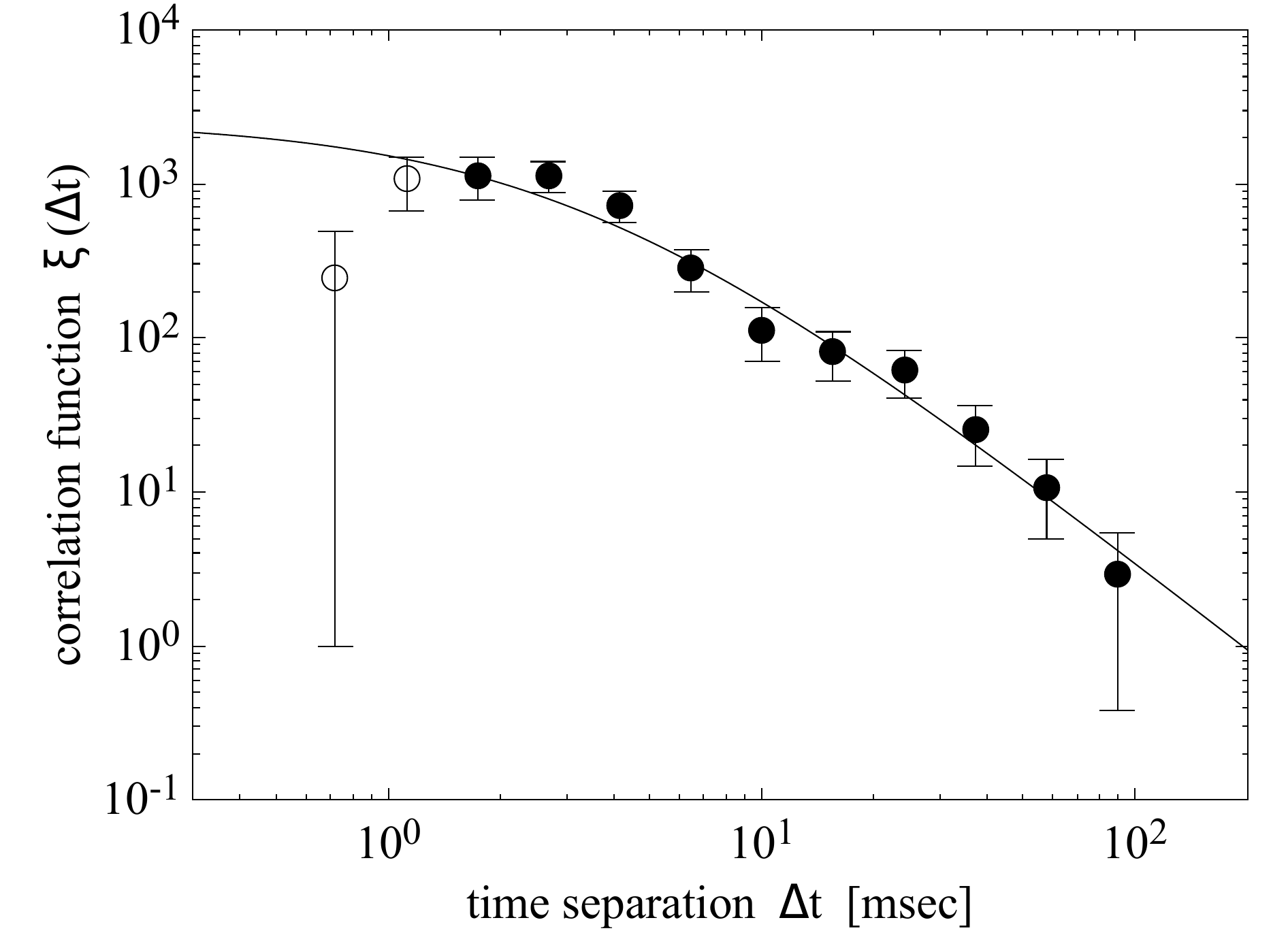}
    \caption{The time correlation function $\xi(\Delta t)$ and model fitting 
    to the radio pulses of SGR 1935+2154 (the R20 sample).}
    \label{fig:chidsqu_fitting}
\end{figure}

Fig. \ref{fig:2Dxi} shows the two-dimensional
correlation function of the R20 sample in the $\Delta t$-$\Delta \mathrm{lg}E$ space. 
No clear signal is detected in
the $\Delta t > 1\,\mathrm{s}$ region, which means 
that radio pulses occur randomly with a certain probability in a given rotation cycle, 
and the occurrence of a pulse in one cycle does not affect the probability of the pulse 
occurring in subsequent cycles. On the other hand, a strong correlation signal
is deteted in the $\Delta t < 1\,\mathrm{s}$ region. The signal shows no dependence
on the energy direction, and the one-dimensional time correlation function
$\xi( \Delta t)$ calculated by pair counts without binning in the energy direction
is shown in Fig. \ref{fig:chidsqu_fitting}. 
Similar to what was seen for extragalactic
FRBs in TT23, we see a correlation signal that decays with 
a power-law and flattening on the shortest time scale comparable to typical pulse width.

We fit this signal with the function of the Omori-Utsu law for earthquakes,
\begin{equation}\label{eq:omori-utsu}
    \xi(\Delta t)=C \, \frac{(\Delta t+\tau)^{-p}}{\tau^{-p}}
\end{equation}
by minimizing $\chi^2$ (Fig.\ref{fig:chidsqu_fitting}). 
The region with the smallest $\Delta t$ is close to the typical pulse width, and
it is susceptible to sub-burst handling because multiple pulses may overlap and
appear to be a single event. Therefore, following TT23, data with $\Delta t$ 
smaller than the peak of $\xi(\Delta t)$ (shown by open symbols in the figure)
were removed from the fit. The best-fit values and
their $1 \sigma$ errors of the model parameters are 
$C = 2.5^{+2.1}_{-1.3} \times 10^3$, $p = 1.9^{+0.8}_{-0.4}$, and 
$\tau = 3.2^{+5.5}_{-2.1}$ msec. 
As expected, the flattening time scale $\tau$ is close to the typical width of a single pulse,
and the power-law correlation signal means that there is no characteristic time scale except for
the duration of one event ($\tau$). 

\begin{figure}
	\includegraphics[width=\columnwidth]{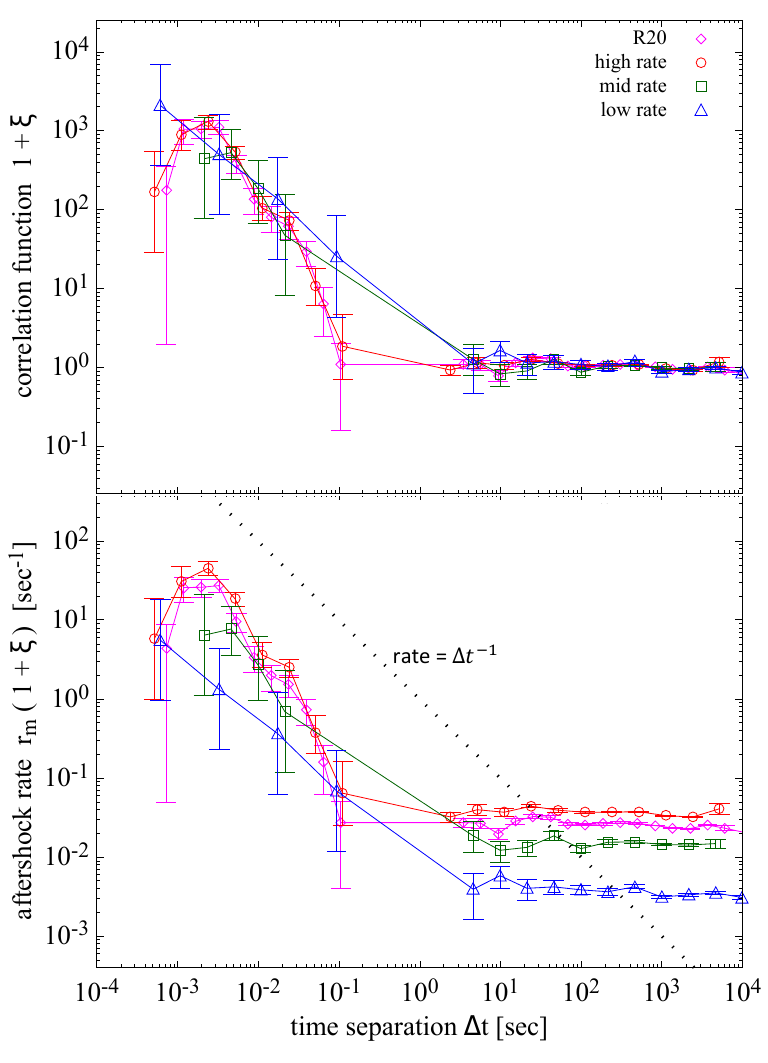}
    \caption{Top: the time correlation functions $1 + \xi(\Delta t)$ for the R20 sample
    and the three sub-samples of high, mid, and low rates. Bottom: the same as the top panel,
    but for the aftershock rate $r_m (1 + \xi)$ following an event, where $r_m$ is the mean
    event rate in a sample.}
    \label{fig:tc_ar}
\end{figure}

In Fig. \ref{fig:tc_ar} we show $(1 + \xi)$ and the aftershock rate 
$r_a \equiv r_m (1 + \xi)$, where
$r_m$ is the mean event rate of the analyzed data set. 
By definition of $\xi$, 
$r_c \equiv r_m \, \xi$ is the occurrence rate of physically correlated aftershocks 
following a single event. In our analysis, a single data set is divided into sub-periods 
where the event rate is assumed to be constant, and event rates for different sub-periods 
may vary widely. In TT23, $r_m$ was estimated by the
mean event rate through all sub-periods 
weighted by the number of pairs, $N_{{\rm pair},i}$
for the $i$-th sub-period, as
\begin{equation}
  r_m = \frac{\sum_i r_{m,i} \, N_{{\rm pair},i}}{\sum_i N_{{\rm pair},i}} \ ,
\end{equation}
where $r_{m,i}$ is the event rate within the $i$-th sub-period.
However, here we use the weight by the number of events, $N_{\rm ev,\it i}$:
\begin{equation}
  r_m = \frac{\sum_i r_{m,i} \, N_{\rm ev, \it i}}{\sum_i N_{{\rm ev}, \it i}}  \ ,
  \label{eq:av_ev}
\end{equation}
because the latter is an appropriate weighting in the sense that $r_a$ is also the average of
those for each subperiod ($r_{a,i}$) weighted by $N_{\rm ev, \it i}$
(see Appendix \ref{appendix:harmonic_mean}).
We applied eq. (\ref{eq:av_ev}) to the TT23 calculation, 
and found that the TT23 results remained virtually unchanged, suggesting that the choice 
between the two averaging methods is not so critical in practice.

Similar to the extragalactic FRBs analyzed in TT23, the correlated aftershock rate
$r_m \, \xi$ is slightly lower than $(\Delta t)^{-1}$, indicating that
the expected number of aftershocks that follow a single event (the branching ratio),
\begin{equation}\label{eq:The branching ratio}
n \equiv \int_0^{\infty} r_m \, \xi(\Delta t) \, d(\Delta t)
\end{equation}
is slightly lower than order unity. Using the best-fit values of $C$, $p$, and $\tau$,
we find $n = 0.18$. Then these results can be interpreted as follows. 
First, a radio pulse occurs, which is the mainshock event occurring
near the specific phase in the rotation cycle for some reason.
Then aftershocks following this mainshock
occur with a probability $n$, and their onset times from the mainshock
follow the Omori-Utsu law, as in the cases of earthquakes and FRBs. 

It is also interesting to see if there are ``aftershocks of aftershocks,'' i.e., 
aftershocks of a similar nature following a single aftershock, 
which can be verified as follows. Of the 464 cycles in which the radio pulses 
were observed, 82 cycles had two or more pulses, and hence the probability 
that an aftershock occurs following the mainshock  
is $82/464 = 0.18 \pm 0.02$ (the error is by $1 \sigma$ Poisson statistics), 
which is consistent with the estimate of the branching ratio $n$.
Of the 82 cycles in which two or more pulses are found, three or more pulses are seen
in 17 cycles, and hence the probability of a third pulse occurring after the second pulse 
is $17/82 = 0.21 \pm 0.06$, which is again consistent with the $n$ value. 
There were no cases where four pulses 
or more occurred in a single cycle, but given that the expected value is $17 \times n = 3.06$, 
it is consistent within about 2 sigma by the Poisson statistics.
These results indeed indicate that aftershocks of aftershocks follow by similar laws,
and are consistent with the concept of the epidemic-type aftershock sequence (ETAS) model \citep{Ogata+1999,Saichev+2006,DEARCANGELIS20161} which does not 
distinguish between mainshocks or aftershocks, but rather assumes that aftershocks occur 
after any event with a common rule. The ETAS model is known to reproduce earthquake data well,
and is also consistent with extragalactic FRBs (TT23). 

The two-dimensional correlation function (Fig. \ref{fig:2Dxi})
shows that the signal is nearly constant in the energy direction, indicating 
a weak or no energy correlation among aftershocks, again similar to the trend seen for FRBs 
and earthquakes in TT23. Another
common characteristic of FRBs and earthquakes found in TT23 was that correlated
aftershock rate $\xi r_m$ remains relatively stable, even as source activity 
(i.e., $r_m$) largely changes. To test whether this holds also for the SGR 1935+2154
radio pulses, we show the time correlation functions for the three sub-samples
of high, mid, and low rates in Fig. \ref{fig:tc_ar}.
Although $\xi r_m$ is not much different between the high- and mid-rate samples, 
that of the low-rate sample is relatively small, which is a different trend
from that found by TT23. This may suggest that when activity is very low, the
probability of aftershocks also decreases, but further research is needed because of the small
statistics in the samples of this study.

\subsubsection{X-ray bursts}

\begin{figure*}
	\includegraphics[width=160mm]{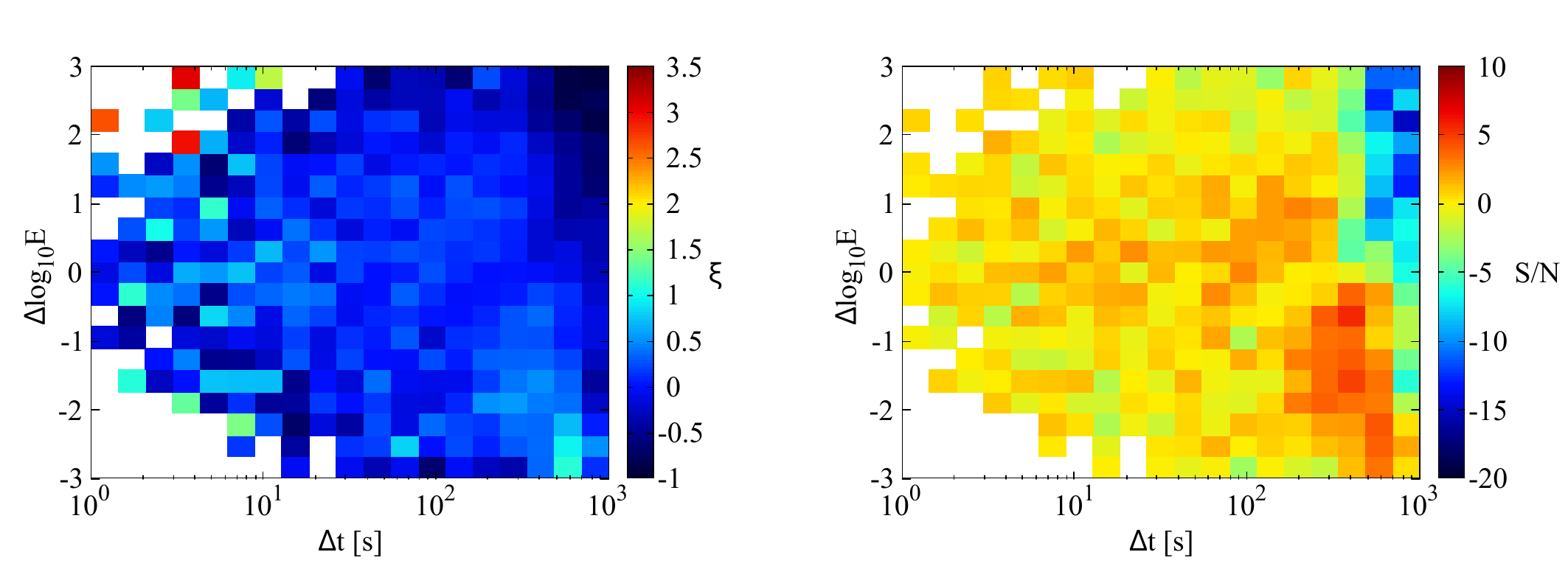}
    \caption{The correlation function $\xi$ (left) and its S/N ratio in the
    2D space of $\Delta t$ and $\Delta \lg E$ for the X20 sample of X-ray bursts
    from SGR 1935+2154. Bins with zero DD pairs are not colored.}
    \label{fig:2Dxi-X20}
\end{figure*}

\begin{figure*}
	\includegraphics[width=160mm]{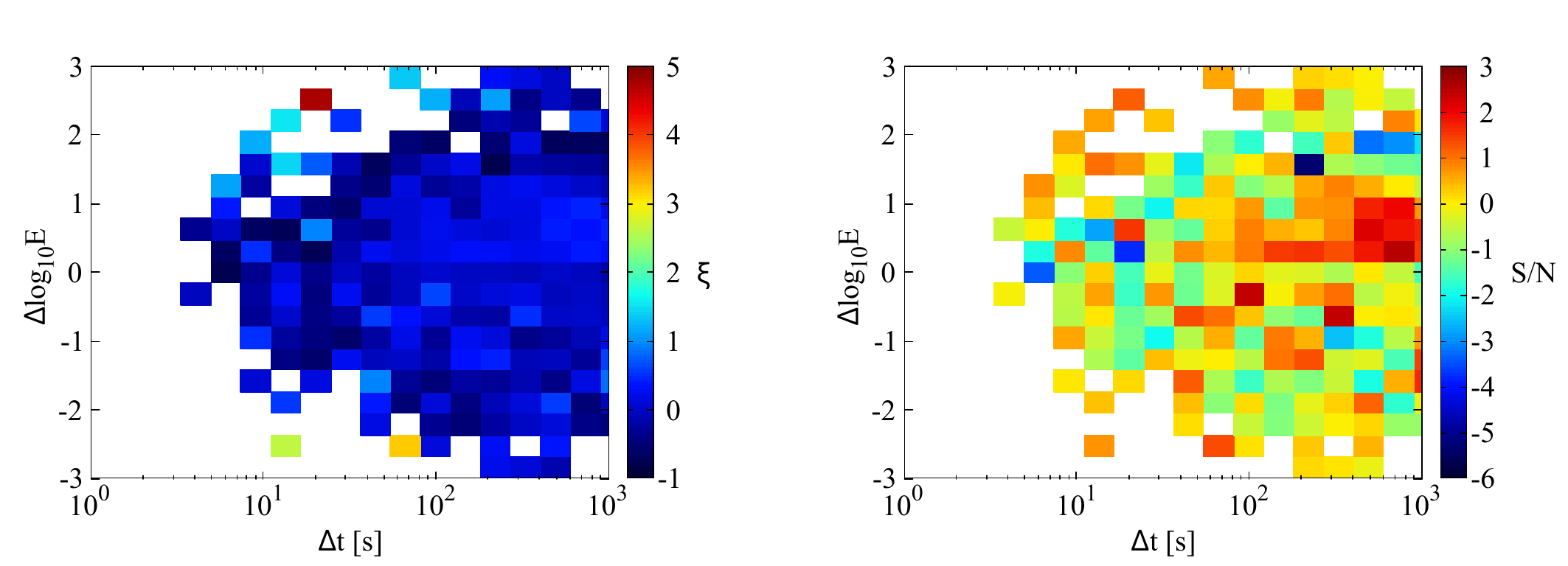}
    \caption{The same as Fig. \ref{fig:2Dxi-X20}, but for the X22 sample. }
    \label{fig:2Dxi-X22}
\end{figure*}

\begin{figure*}
	\includegraphics[width=160mm]{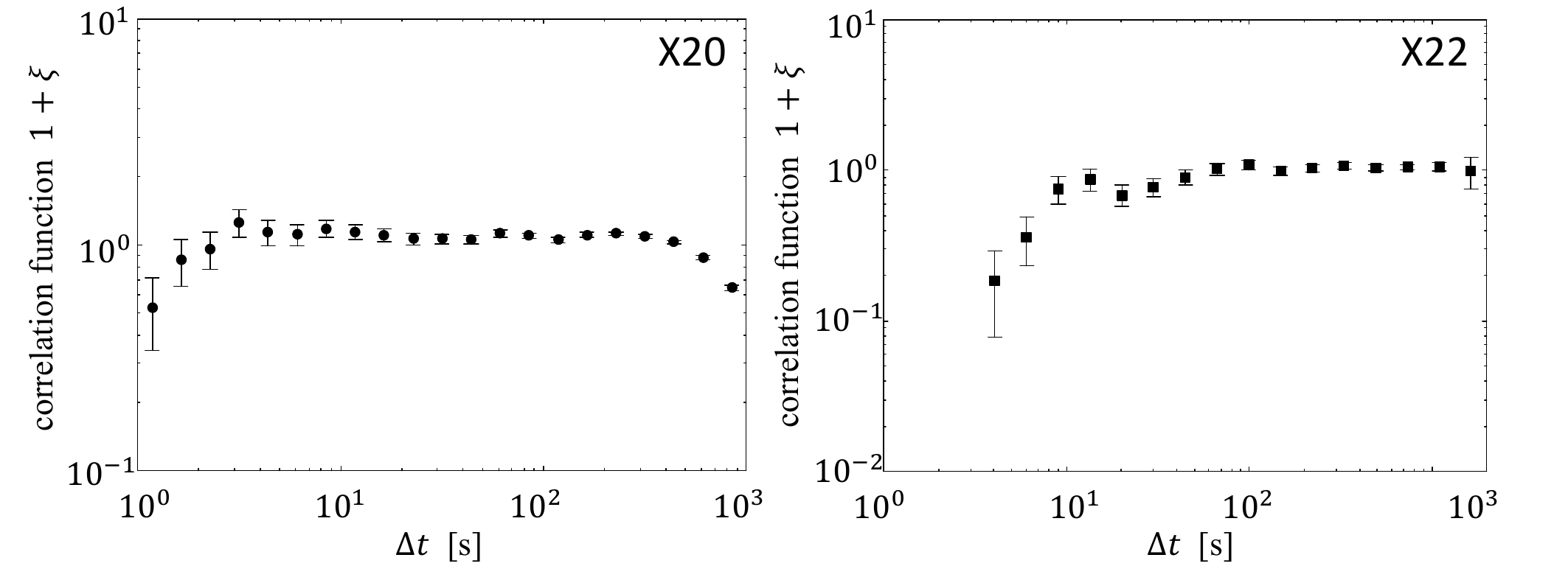}
    \caption{The time correlation function $1 + \xi(\Delta t)$ for the X20 (left)
    and X22 (right) samples of the X-ray bursts from SGR 1935+2154.}
    \label{fig:1Dxi-X}
\end{figure*}

The computed 2D correlation functions for the X20 and X22 data sets are shown in 
Figs. \ref{fig:2Dxi-X20} and \ref{fig:2Dxi-X22}, respectively,
and the 1D time correlation functions are shown in Figure \ref{fig:1Dxi-X}.
The values of the correlation function $\xi$ are close to zero in most regions, 
and no strong correlation signals as seen in the radio pulses were detected.
In both data sets, $\xi$ is negative in the region $\Delta t < 10$ s, 
i.e., an anti-correlation signal is visible. This time scale is close to the typical 
duration of an X-ray burst \citep{younes2020nicer}, and the anti-correlation is
likely by an effect of multiple overlapping bursts that are not counted as two 
independent events. It should be noted that $65\%$ of bursts in the X20 sample
show multi-peaked structures \citep{younes2020nicer}. 

On the other hand, 
non-zero $\xi$ signals with significantly large signal-to-noise (S/N) ratios are visible 
on long-time scales of $\Delta t >$ 100 s. However, this region has many pairs,
and slight systematic errors can easily produce a false signal.
In particular, if the event rate or energy distribution of the bursts changes over 
a long time, there is a risk of false correlation signals compared to the random sample, 
where they are assumed to be constant.  In conclusion, no significant time/energy
correlations were detected in the X-ray bursts, though the time scales of the correlated
radio pulses cannot be examined because they are hidden by the long duration of X-ray burst.

\section{Interpretation}

\subsection{The origin of pulses/bursts from SGR 1935+2154}

The most notable result of our analysis is that the radio pulses of SGR 1935+2154 are 
accompanied by aftershocks of a similar nature to those of extragalactic FRBs and earthquakes,
satisfying four of 
the five similarity points between FRBs and earthquakes listed in Section
\ref{section:intro}. The only difference between FRBs and earthquakes is the value of 
the Omori-Utsu index ($p \sim 2$ for FRBs and $\sim 1$ for earthquakes), 
and the value of SGR 1935+2154 is close to that of FRBs.
In addition to the similarity between FRBs and SGR 1935+2154 radio pulses
in the characteristics of the radio signals
(narrow-band emissions and frequency drifts) already reported by 
\citet{zhu+23,wang2023atypical}, 
the results of this work strongly suggest that these two 
are physically similar phenomena.

However, there is one significant difference between FRBs and SGR 1935+2154;
the SGR 1935+2154 radio pulses are concentrated in a specific phase of the rotation cycle.
This can be interpreted as the first event in a cycle being triggered 
with a probability of 2.5\% per cycle by some external force
synchronized with the rotation of the neutron star, and the following aftershocks 
within the same cycle
being similar in nature to those of FRBs or earthquakes.
The average single pulse luminosity is about 5 orders of magnitude smaller than the spin-down
luminosity ($4.3 \times 10^{34}$ erg/s), and the energy emitted as radio pulses during an
observation time is 10 orders of magnitude smaller than the spin-down energy \citep{zhu+23}.
Therefore, the rotational energy of the star is sufficient to explain the radio pulse.

Given the similarity to earthquakes and the popular theoretical
concept linking starquakes and magnetar activities, 
one promising hypothesis is that they are related to breakup 
or cracking of the neutron star crust. In this case,
an interesting question is what is the periodic external force that triggers the 
first radio pulse in one rotation cycle. An isolated neutron star rotating in a vacuum 
is not expected to feel periodically varying external forces or torques.
In binary systems, tidal forces from the companion star may be periodic external forces, 
but no signs of binarity have been reported for SGR 1935+2154 \citep{Lyman_2022,Chrimes_2022}.
A noteworthy observation here is that radio pulses of magnetars are generally transient and 
are observed in association with an X-ray outburst \citep{kaspi2017magnetars,esposito2018magnetars}.
During the active period including an outburst, there is likely to be mass ejection from 
the magnetar and hence relatively dense gas is thought to exist in the vicinity 
of the neutron star magnetosphere compared to low activity phases.
Interestingly, NICER observations of SGR 1935+2154 \citep{Younes+23} show that a large spin-down 
glitch occurred three days 
before FAST detected three moderately bright FRB-like radio bursts \citep{good+20,Pleunis+20}, and 
then the periodic radio pulses were detected within one day from the FRB-like bursts. 
This suggests that there was a decrease in the stellar angular momentum associated with 
a mass loss before the emergence of the periodic radio pulses.
Then non-axisymmetric magnetic interactions between such surrounding material and 
the central star may produce torques that vary with the rotational period of the star.
Whether this mechanism causes the first pulse within a cycle to occur at precisely a particular 
phase requires a more detailed study, and we encourage theoretical studies on this direction. 

A mechanism commonly considered to explain a pulse with a short time width at a 
particular rotational phase 
is that the pulse is visible when the collimated radiation crosses the line of sight.
The radio pulses from SGR 1935+2154 are detected from only 2.5\% of rotation cycles.
Such low duty cycles and multiple pulses within one cycle have been observed from other pulsars 
and are commonly explained by considering patchy beams.
A potential problem with this model, however, is whether it can naturally account for the 
systematic aftershock law observed in the delay time distribution of radio pulses from 
SGR 1935+2154. Theoretical investigation about this issue 
is also an interesting future research direction.

We did not detect any significant time correlation signals for the X-ray bursts.
However, the time scale of the correlations observed in radio pulses is less than 100 msec, 
which is shorter than the typical duration of X-ray bursts
($\sim$ 0.1--10 s, \citealt{younes2020nicer}).
Therefore X-ray bursts may have correlations similar to those of 
radio pulses, but they are not detected because of 
the superposition of radiation from successive bursts.
On the other hand, X-ray bursts may be essentially a different phenomenon 
from radio pulses, given the following observed facts. 
The peak time of X-ray bursts occurs at random phases in the rotation cycle
\citep{younes2020nicer}. The luminosity of X-ray bursts ($10^{37-39}$ erg/s) 
exceeds the spin-down luminosity, as does
the average energy release rate estimated from the total energy of all bursts 
($4.8 \times 10^{40}$ erg) that occurred during the 1120 s observation time
of the X20 sample \citep{younes2020nicer}. Therefore, X-ray bursts cannot be explained by 
rotational energy, but are most likely produced by magnetic energy of the magnetar,
which is in contrast to radio pulses synchronized with rotation.
However, much brighter FRB-like radio bursts observed from SGR 1935+2154 may have
a common physical origin with X-ray bursts, as discussed next.

\subsection{Discussion on the connection with FRBs}

Although we have argued that both FRBs and the radio pulses of SGR 1935+2154 are 
likely related to the crustal breakup of neutron stars,
there is a significant difference 
in energy between FRB-like bursts ($\sim 10^{34-35}\,\mathrm{erg}$ for FRB 20200428,
\citealt{andersen2020bright,bochenek2020fast})
and radio pulses ($\sim10^{26-28}\,\mathrm{erg}$,
\citealt{zhu+23,wang2023atypical}). 
The peak radio luminosity of FRB 20200428 ($\sim 3 \times 10^{36}$ erg/s, 
\citealt{andersen2020bright})
exceeds the spin-down luminosity, 
which is difficult to explain by spin-down unless rotational energy is somehow stored in 
a certain time and released all at once. 
Furthermore, the phases of the FRB-like bright radio bursts from SGR 1935+2154 observed 
so far do not match that of the periodic radio pulses \citep{zhu+23}.
These facts suggest that the FRB-like bursts of SGR 1935+2154 and even brighter extragalactic 
FRBs are driven by a completely different energy source than rotational energy, 
most likely magnetic energy.

These facts imply that even if the energy sources and scales are completely different, 
radio bursts with remarkably similar radiation properties and aftershock nature can occur.
The similarity may be attributed to physical processes such as starquakes 
rather than energy sources.
This then suggests that there may be yet other populations of FRB-like radio burst 
phenomena driven by an energy source that is neither magnetic nor rotational.
For example, radio bursts may occur as tidal forces induce starquakes
in close encounters involving a neutron star in globular clusters,
which may be the case for the repeating FRB source in a globular cluster
of the nearby galaxy M81 \citep{kirsten+22} . 
It is important to search for short-duration radio bursts from more magnetars and non-magnetar 
neutron star populations to investigate their diversity and commonality.

\section{Conclusion}

In order to investigate the time and energy correlation properties of burst phenomena
of SGR 1935+2154, the Galactic magnetar that is known to produce FRBs, we analyzed the 
two-point correlation 
functions in time and energy space for the periodic radio pulses (563 pulses)
and X-ray short bursts (two data sets, each containing 205 and 374 events).

Radio pulses occur at a particular phase of the rotational cycle,
but randomly with a probability of 2.5\% of all the cycles. 
By examining time correlations for multiple pulses within a single 
cycle, we found that their aftershock properties are similar to those of FRBs and earthquakes
in four of the five similarities reported in TT23. 
Namely, (1) each pulse is accompanied by an aftershock with a probability 
of about $n \sim 20\%$, 
(2) its distribution of the delay time $\Delta t$ follows the Omori-Utsu law, 
$\propto (\Delta t + \tau)^{-p}$ with $p \sim 2$, 
(3) the flattening time scale $\tau$ is close to the typical pulse width (1--10 msec), 
and (4) there is no strong correlation about energy of aftershock pulses.
In this picture, there is no distinction between mainshocks and aftershocks, 
and aftershocks of aftershocks occur with the same statistical properties. 
This is confirmed by the ratio of the number of cases in which two and three pulses occurred in 
a single rotation cycle. This is consistent with the picture of the ETAS model, 
which is known to be in good agreement with the earthquake data.
Regarding another similarity reported in TT23, namely that the aftershock rate is 
constant regardless of source activity, the SGR 1935+2154 radio pulses may differ at this point, 
since the aftershock rate is lower in
the low-rate sample, during which the radio pulse detection rate was about 10 times
lower than other sub-samples. 
However, the present sample lacks sufficient statistics to conclude this unequivocally, 
and further data and study are needed.

Thus, a natural interpretation of these results would be to assume that periodic radio pulses are
due to seismic ruptures in the surface crust of the neutron star,
though we cannot rule out the possibility that an entirely different physical process 
than starquakes could cause aftershocks of similar properties. 
In this scenario, the power-law time correlation function requires that the 
time scale of crustal ruptures is less than msec, and this should provide constraints on the physical 
properties of the neutron star crust (e.g., the propagation speed of seismic waves) 
and the size of rupture regions,
which can be verified by future theoretical studies.
Furthermore, the pulses occur around the specific phase within the 3.24-second period, 
suggesting that the initial pulse in a cycle is triggered by periodically
varying external force on the crust, followed by accompanying aftershocks similar to FRBs and 
earthquakes. 
The energy emitted as periodic radio pulses is sufficiently small to be
explained by the spin-down luminosity of the star.
There is no sign of a binary system for SGR 1935+2154, and magnetar 
radio pulses are generally transient and appear in association 
with outbursts. Therefore we speculate that this periodic 
external force may be the interaction between the rotating magnetosphere and 
dense surrounding material 
ejected by an outburst, rather than tidal forces from the companion star.
Another possibility is that the short-width pulses originate from beaming of radiation 
crossing the line of sight, and for some reason they are followed by pulses with
a power-law distribution of delay times. 

For X-ray short bursts, on the other hand, we could not find any significant correlated 
signals in the $\Delta t \gtrsim 1$ s region.
Although we cannot rule out the possibility that the correlation of 
$\Delta t \lesssim 100$ ms found in radio pulses are masked by the longer duration
($\sim 1$ s) of X-ray bursts, it is also possible that X-ray bursts are essentially a different
phenomenon from periodic radio pulses. The phase of X-ray bursts is random relative to 
the rotation cycle, and both the luminosity of individual bursts and the burst energy 
generation rate averaged over an observation period exceed the spin-down luminosity.
Thus, unlike periodic radio pulses, the energy source of X-ray bursts cannot be stellar rotation.

It is surprising that periodic radio pulses of SGR 1935+2154
exhibit aftershock properties similar to those 
of extragalactic FRBs, even though the radio pulses are $\sim 10^7$ times
fainter than FRB 20200428.
Since the luminosity of the FRB-like bursts in SGR 1935+2154 exceeds the spin-down luminosity 
and the phases at which they occur are random relative to the rotational cycle, 
stellar rotation is unlikely to be the energy source for these bursts and the even brighter
extragalactic FRBs. Then, the similarity in the nature of radio emissions and aftershocks 
between FRBs and the magnetar periodic radio pulses should be attributed 
not to energy sources but to physical processes such as crustal rupture.
This hypothesis suggests that the common essence of the FRB phenomenon 
is not what the energy source is, but starquake processes in neutron stars.
FRB-like events could then occur in neutron stars with various properties 
and environments, and
it is important to search for FRB-like radio bursts of various energy scales from neutron 
stars of various populations including non-magnetars.
It would also be interesting to investigate whether a series of radio pulses from other magnetars
or ordinary part-time pulsars also show a similar time correlation to SGR 1953+2154.

\section*{Acknowledgements}
TT was supported by the JSPS/MEXT KAKENHI Grant Number 18K03692.

\section*{Data Availability}

The data newly derived in this article (e.g. correlation function values) 
will be shared on reasonable request to the corresponding author.



\bibliographystyle{mnras}
\bibliography{Similarity} 




\appendix

\section{Optimal calculation of the mean event rate for all sub-periods}
\label{appendix:harmonic_mean}

After the correlation function $\xi$ has been computed for the entire data set, 
the mean event rate $r_m$ is needed to 
compute the aftershock rate $r_a(\Delta t) \equiv [1 + \xi(\Delta t)] \, r_m$. 
In our calculations, one entire data set is divided into many sub-periods, and
it is not obvious how best to calculate the mean $r_m$ for the entire data set 
when the event rate and sub-period length vary widely among sub-periods.
We show here that it is appropriate to take an average of $r_m$ 
weighted by the number of events in each sub-period.

Consider an average of the aftershock rates over all sub-periods with some weight $w_i$, 
\begin{eqnarray}
  \bar{r}_a(\Delta t)  \equiv \frac{\sum_i w_i \, r_{a,i}}{\sum_i w_i} \ ,
\label{eq:r_a_nobias}
\end{eqnarray}
where
\begin{eqnarray}
  r_{a,i}(\Delta t) = (1 + \xi_i) \, r_{m,i} = \frac{DD_i}{RR_i} \, r_{m,i}
\end{eqnarray}
is the aftershock rate in the $i$-th sub-period, and the subscript $i$ indicates that 
it is the quantity within that sub-period. 
If we take $w_i = RR_i / r_{m,i}$ as the weight here, 
$\bar{r}_a$ is calculated as
\begin{eqnarray}
 \bar{r}_a(\Delta t) &=& \frac{\sum_i DD_i}{\sum_i RR_i / r_{m,i}}
 = \frac{\sum_i DD_i}{\sum_i RR_i} \frac{\sum_i RR_i}{\sum_i RR_i / r_{m,i} }
  \nonumber \\
 &=& (1 + \xi) \left( \frac{\sum_i r_{m,i}^{-1} \, RR_i}{\sum_i RR_i} \right)^{-1} \ ,
\end{eqnarray}
and $\bar r_a$ and $r_a$ become the same if we define $r_m$ as
\begin{eqnarray}
  r_m \equiv \left( \frac{\sum_i r_{m,i}^{-1} \, RR_i}{\sum_i RR_i} \right)^{-1}
\end{eqnarray}
(the harmonic mean of $r_{m,i}$ weighted by $RR_i$). 

The number of random pairs $RR_i$ is proportional to $r_{m,i}^2$ and 
to the length of sup-period $T_i$ if the $\Delta t$ of interest is much shorter 
than the typical sub-period length. Therefore, the weight $w_i$ employed
here for $r_a$ is actually $w_i \propto r_{m,i} \, T_i = N_{\rm ev, \it i}$,
where $N_{\rm ev, \it i}$ is the event number in the
$i$-th sub-period. Then it can be shown that the above expression of
$r_m$ is also an arithmetic mean of $r_{m,i}$ with the weight by $N_{\rm ev, \it i}$, as
\begin{eqnarray}
  r_m = \frac{\sum_i r_{m,i}^2 \, T_i}{\sum_i r_{m,i} \, T_i} = 
  \frac{\sum_i r_{m,i} \, N_{\rm ev, \it i}}{\sum_i N_{\rm ev, \it i}} \ .
\end{eqnarray}
Therefore, the mean event rate $r_m$ weighted by $N_{\rm ev, \it i}$
in each sub-period is appropriate for determining the aftershock rate
during the entire period by $r_a = (1 + \xi) \ r_m$,
because $r_a$ in this case
becomes the average of $r_{a,i}$ also weighted by $N_{\rm ev, \it i}$.


\bsp	
\label{lastpage}
\end{document}